*Reply to comment by Yermolaev & Lodkina on "Solar Wind and Heavy Ion Properties of Interplanetary Coronal Mass Ejections" by Owens (2018) in arXiv:2008.05160*

*Mathew Owens*

Here I reply to the comment by Yermolaev & Lodkina [published on arXiv on 12/08/2020 https://arxiv.org/abs/2008.05160] on "Solar Wind and Heavy Ion Properties of Interplanetary Coronal Mass Ejections" by Owens (2018).

Before I get in to the specifics, I'd like to make two points. Firstly, if arXiv is to be used as a means to make unrefereed Comments on published articles, the authors should at least be notified by email so that they have the opportunity to respond. Secondly, I'd like to note for the record that an extremely similar Comment article, with some identical phrasing, was made by Yermolaev & Lodkina on "A Statistical Study of the Plasma and Composition Distribution inside Magnetic Clouds: 1998–2011" by J. Huang et al. (2020) [**https://tinyurl.com/y6qq3lbr**].

The Yermolaev & Lodkina comment on Owens (2018) makes three criticisms, summarised in italics, which are responded to in turn:

1. *"The article by Owens (2018) did not reference the article by Lynch et al. (2003)"*

    This is accurate. I was not aware of the Lynch study at the time (and neither apparently was the referee of Owens, 2018). It is unfortunate, as Lynch et al. (2003) is both an interesting and relevant study. (Apologies, Ben: while I try to be thorough in reviewing the literature, this kind of oversight is inevitable from time to time.)

    However, it should be noted that Lynch et al. (2003) dealt with a relatively small sample of magnetic clouds, whereas Owens (2018) considered the ambient solar wind and a larger number of both magnetic cloud and non-cloud ICMEs, with the specific aim of comparing and contrasting the three populations. The methods of analysis between Lynch et al (2003) and Owens (2018) also differ substantially. Thus had I been aware of Lynch et al. (2003) at the time of writing, I would certainly have cited it and made comparison, but my study and conclusions would have been largely unchanged from the published form.

2. *"The results obtained in work by Owens (2018) either partially or completely contradict those previously obtained regarding the mass and charge composition of plasma in magnetic clouds… which indicates the incorrectness of the event selection, data analysis and conclusions presented"*

    As detailed in Owens (2018), event selection was made using an existing, refereed event catalogue, namely the updated version of the Cane & Richardson (2003) ICME catalogue, available here: http://www.srl.caltech.edu/ACE/ASC/DATA/level3/icmetable2.htm. This is something of an industry standard, with approximately 1000 citations across the 2003 and 2010 papers. If Yermolaev & Lodkina have issues with this event catalogue, their criticisms should be addressed to Ian Richardson and Hilary Cane.

The comment then implies Owens (2018) does not find enhanced alpha-to-proton, O7-to-O6 and Fe-charge-state ratios within magnetic clouds, which should be expected on the basis of previous results. This is not correct: Owens (2018) show magnetic clouds to be significantly enhanced in all three of these properties compared with both the ambient solar wind and non-magnetic cloud ICMEs. See Figures 5, 7 and 8. While the mean values for two of the three of these properties (alpha-to-proton and O7-to-O6 ratios) are not quite as high as those quoted in the comment, this is to be expected for two reasons. Firstly, the quoted numbers are at the high end of the ranges and were not originally reported as minimum selection criteria. Secondly, those studies considered relatively small samples of magnetic clouds, likely selected for their exceptionally clear and unambiguous signatures. On average, values are likely to be lower when less "pristine" events are included in the sample.

3. "*The work [Owens 2018] uses the term "1-sigma variations", but does not explain what the value means or how it was calculated*"

In Figures 3 and 4, Owens (2018) show uncertainty using the 1-, 2- and 3-sigma ranges. In Figure 5 only the 1-sigma range is shown, for clarity. In Figures 6 and 7 the full distribution functions are presented. The text explains that the 1-sigma range spans 1-sigma of the distribution. While more detail could have been provided, this is a standard statistical method and I (and the referee) assumed most researchers would be familiar with it (e.g., it is also commonplace in our field to use terms like median or standard deviation without further explanation).

Finally, Yermolaev & Lodkina conclude that "*our assessment of data presented in Owens (2018) indicates that the work does not meet basic requirements of scientific ethics*". An identical claim was made at Huang et al (2020) on the basis of similarly minor criticisms. Even if one agrees with the three points presented by Yermolaev & Lodkina, it is difficult to see how they constitute a breach of basic ethics. Thus the accusation is noted here but does not merit further discussion.